\documentclass[conference]{IEEEtran}
\IEEEoverridecommandlockouts
\usepackage{amsmath}
\usepackage{amssymb}
\usepackage{amsfonts}
\usepackage{algorithm}
\usepackage{dsfont}
\usepackage{float}
\usepackage{cite}
\usepackage{graphicx}
\usepackage{epsfig}
\usepackage{subfigure}
\usepackage{psfrag}
\usepackage{xcolor}
\usepackage{url}
\usepackage[colorlinks,linkcolor=black,urlcolor=black,anchorcolor=black,citecolor=black,hyperfootnotes=true]{hyperref}
\def\BibTeX{{\rm B\kern-.05em{\sc i\kern-.025em b}\kern-.08em
    T\kern-.1667em\lower.7ex\hbox{E}\kern-.125emX}}

\begin{document}
\title{A Two-Stage Radar Sensing Approach based on MIMO-OFDM Technology}
\author{\IEEEauthorblockN{Liang Liu and Shuowen Zhang}
	\IEEEauthorblockA{EIE Department, The Hong Kong Polytechnic University. \\Email: liang-eie.liu@polyu.edu.hk, shuowen.zhang@polyu.edu.hk}
}

\maketitle

\begin{abstract}
Recently, integrating the communication and sensing functions into a common network has attracted a great amount of attention. This paper considers the advanced signal processing techniques for enabling the radar to sense the environment via the communication signals. Since the technologies of orthogonal frequency division multiplexing (OFDM) and multiple-input multiple-output (MIMO) are widely used in the legacy cellular systems, this paper proposes a two-stage signal processing approach for radar sensing in an MIMO-OFDM system, where the scattered channels caused by various targets are estimated in the first stage, and the location information of the targets is then extracted from their scattered channels in the second stage. Specifically, based on the observations that radar sensing is similar to multi-path communication in the sense that different targets scatter the signal sent by the radar transmitter to the radar receiver with various delay, and that the number of scatters is limited, we show that the OFDM-based channel training approach together with the compressed sensing technique can be utilized to estimate the scattered channels efficiently in Stage I. Moreover, to tackle the challenge arising from range resolution for sensing the location of closely spaced targets, we show that the MIMO radar technique can be leveraged in Stage II such that the radar has sufficient spatial samples to even detect the targets in close proximity based on their scattered channels. Last, numerical examples are provided to show the effectiveness of our proposed sensing approach which merely relies on the existing MIMO-OFDM communication techniques.
\end{abstract}
%
%

\newtheorem{definition}{Definition}
\newtheorem{assumption}{Assumption}
\newtheorem{lemma}{\underline{Lemma}}
\newtheorem{example}{Example}
\newtheorem{theorem}{Theorem}
\newtheorem{proposition}{Proposition}
\newtheorem{conjecture}{Conjecture}
\newtheorem{remark}{Remark}
\newcommand{\mv}[1]{\mbox{\boldmath{$ #1 $}}}

\section{Introduction}\label{sec:Introduction}

Radar and wireless communication are probably the two most successful applications of radio technology over the past decades. Recently, there is growing interests in integrating the functions of sensing the environment and conveying the information in a common system via reusing the same radio frequency (RF) signals \cite{sturm2011waveform,zheng2019radar,paul2016survey,liu2020joint}. For example, the intelligent transportation industry, which requires both strong sensing capability to detect adjacent vehicles and communication capability to share these information, can take advantage of such an integrated system for reducing the number of antennas equipped at the smart vehicles \cite{kumari2017ieee}. Further, the performance of communication can be improved by sensing as well. For instance, the sensing information can be leveraged for beam selection and alignment in millimeter wave systems \cite{muns2019beam}.


Despite the promising benefits, a big obstacle exists for truly realizing a joint communication and radar system: the different philosophy to design the RF waveforms \cite{sturm2011waveform}. Specifically, for the radar systems, the RF waveforms are deterministic and need to satisfy some autocorrelation properties such that the receiver is able to sense the high dynamic range of targets when applying the correlation processing. On the other hand, because information is represented by randomness, the modulated signals for communication have to change dynamically in amplitude, frequency, or phase, without preserving the autocorrelation properties. For unifying the communication and sensing technologies, there are thus two possible approaches: embedding the information onto the radar signals with good autocorrelation properties \cite{kumari2017ieee}, and sensing the targets based on communication signals via new signal precessing techniques other than correlation processing \cite{sturm2009novel}. While the former approach is appealing for applications requiring low data rate, the latter one seems to be the only option for moderate or high data rate applications. Therefore, this paper focuses on the advanced signal processing techniques to enable the radar sensing function using the communication RF signals.

In particular, this paper adopts the orthogonal frequency division multiplexing (OFDM) technique for radar sensing, the motivation of which is two-fold. First, OFDM is widely used in the 4G and 5G cellular systems, and from the compatibility perspective, it is necessary to investigate how to sense the targets in OFDM systems. Second, radar sensing resembles multi-path communication in the sense that multiple targets scatter the signal from the transmitter to the receiver with various delay. Since the OFDM technique is effective to combat the inter-symbol interference for the multi-path communication, it is reasonable to believe in its power for radar sensing as well. To harvest the advantages of the OFDM technique in the multi-path scenario, this paper proposes a two-stage signal processing approach to sense the targets based on the OFDM signals scattered by the targets to the radar receiver. In the first stage, the radar estimates the channel of each path from the radar transmitter to one particular target to the radar receiver, which is determined by the distance and angle from this target to the radar. In the second phase, the radar estimates the location of each target based on its scattered channel. One challenge of the above scheme lies in the range resolution \cite{Richards10}: if two targets are too close, their scattered signals may arrive at the radar receiver within the same OFDM sample period. In this case, the radar can merely estimate the overall channel, instead of individual channels, of these two targets, and thus may not accurately estimate their location. To combat the above challenge, we adopt the multiple-input multiple-output (MIMO) radar technique \cite{Li07,Haimovich08} such that the radar has sufficient spatial samples to resolve the location of targets with similar range to it in Stage II.

\section{System Model}
\subsection{General System Model}

We consider an MIMO-OFDM based joint sensing and communication system, where a transmitter is equipped with a mono-static radar and makes dual use of the RF signals for conveying information to the information receivers and detecting the targets based on their scattered signals at the same time. Since the information transmission techniques are already very mature, in the rest of this paper, we mainly focus on how to detect the targets via utilizing the information-bearing RF signals. We assume that the radar is equipped with $M_T$ transmit antennas for sending the RF signals and $M_R$ receive antennas for detecting the signals scattered from the targets. Moreover, let $d_{{\rm max}}$ meters (m) denote the maximum radar detection range, i.e., only the scattered signals from the targets whose distance to the radar is within $d_{{\rm max}}$ m are sufficiently strong for detection. Further, let $0<\theta_{{\rm max}}\leq 2\pi$ denote maximum angle that can be detected by the radar. In the above detectable region, define $K$ as the number of targets, while $d_k<d_{{\rm max}}$ m and $\theta_k\in (0,\theta_{{\rm max}}]$ as the distance and angle from target $k$ to the radar, respectively, $k=1,\ldots,K$.

For the OFDM technique, we assume that there are $N$ sub-carriers in the system, and the sub-carrier spacing is denoted by $\Delta f$ Hz. As a result, the sampling rate for the radar OFDM signal is $F_s=N\Delta f$ Hz. Then, the range resolution in terms of m is defined as \cite{Richards10}
\begin{align}\label{eqn:range resolution}
\Delta d=\frac{c_0}{2F_s}=\frac{c_0}{2N \Delta f},
\end{align}where $c_0$ m/s denotes the speed of light. For instance, in an OFDM system where $N=1024$ and $\Delta f=15$ KHz, the range resolution is $\Delta d\approx 9.77$ m, which is quite large in practice.

Note that two targets $k$ and $k'$ with $|d_k-d_{k'}|\leq \Delta d$ m are hard to be resolved in range because their scattered signals arrive at the radar receiver within the same sample period. Particularly, for a radar transmit signal, its scattered signal from target $k$ will cause a delay of
\begin{align}
\tau_k=\left\lceil\frac{2N\Delta f d_k}{c_0}\right\rceil, ~~~ k=1,\ldots,K,
\end{align}OFDM samples at the radar receiver side. Then, we can define
\begin{align}
 & \Omega_l\hspace{-2pt}=\hspace{-2pt}\{k_l\hspace{-2pt}:\hspace{-2pt}\tau_{k_l} =l\}\hspace{-2pt}=\hspace{-2pt}\left\{k_l\hspace{-2pt}:\hspace{-2pt}\frac{lc_0}{2N\Delta f}<d_{k_l}\leq \frac{(l+1)c_0}{2N\Delta f}\right\}, \label{eqn:target1} \\
 & K_l\hspace{-2pt}=\hspace{-2pt}|\Omega_l|\geq 0, ~~~ l=1,\cdots,L_{{\rm max}}, \label{eqn:number of targets}
\end{align}as the range cluster of all the targets whose scattered signals are delayed by $l$ OFDM samples at the radar receiver side and the number of targets in this cluster, respectively, where
\begin{align}\label{eqn:maximum deley}
L_{{\rm max}}=\left\lceil \frac{2d_{{\rm max}} N\Delta f}{c_0} \right\rceil,
\end{align}is the maximum delay in terms of samples caused by a target at the distance of $d_{{\rm max}}$ m. It is observed from (\ref{eqn:target1}) that the definition of range cluster is related to the range resolution defined in (\ref{eqn:range resolution}). For convenience, after clustering, if the $k$th target in the radar system is the $i$th target in cluster $l$, then we also use $D_{l,i}=d_k$ m and $\Theta_{l,i}=\theta_k$ to denote the distance and angle from this target to the radar, respectively.

\begin{figure}[t]
  \centering
  \includegraphics[width=8.5cm]{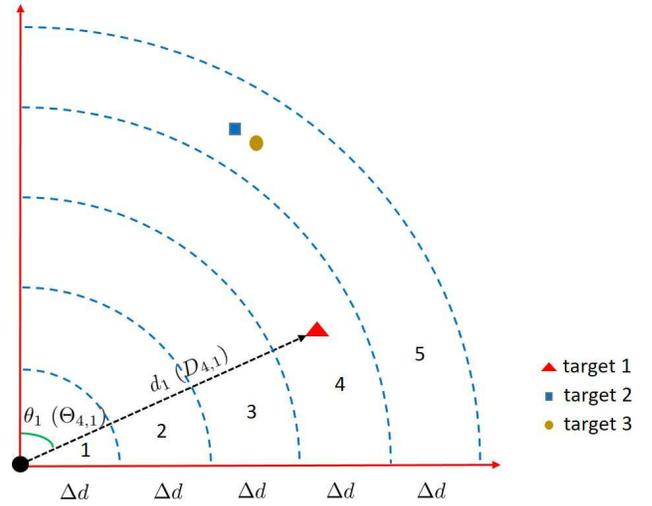}
  \caption{An illustration of the radar system: there are $K=3$ targets located in $L_{{\rm max}}=5$ clusters, while each cluster is spaced by the range resolution, i.e., $\Delta d$ m, and its maximum angle is $\theta_{{\rm max}}=\pi/2$. In this system, no target is in clusters 1, 2, and 3, i.e., $\Omega_1=\Omega_2=\Omega_3=\emptyset$, target 1 is in cluster 4, i.e., $\Omega_4=\{1\}$, and targets 2 and 3 are in cluster 5, i.e., $\Omega_5=\{2,3\}$. Moreover, for target 1 (denoted by the red triangle), its distance and angle to the radar are depicted as $d_1$ or $D_{4,1}$ m and $\theta_1$ or $\Theta_{4,1}$ (since this target is the 1st target in cluster 4).}\label{Fig1}\vspace{-15pt}
\end{figure}

An example of the above radar system is shown in Fig. \ref{Fig1}, where $d_{{\rm max}}=5\Delta d$ and $\theta_{{\rm max}}=\pi/2$.

\subsection{Radar Sensing Signal Model}\label{sec:Signal Model}
Suppose that there is no obstacle between the radar and targets. Then, the scattered channel from the radar transmitter to each target to the radar receiver merely depends on the target's location. Specifically, given any $1\leq m_T\leq M_T$ and $1\leq m_R\leq M_R$, let $g_{m_R,m_T}(d_k,\theta_k)$ denote the channel from transmit antenna $m_T$ to target $k$ to receive antenna $m_R$, which is a function of $d_k$ and $\theta_k$. For example, if the radar employs a uniform linear array (ULA) where the antenna spacing is $d_{{\rm A}}$ m and the operating frequency of the OFDM system is much larger than the total bandwidth such that the signal wavelength in all sub-carriers can approximately be considered identical, then the scattered channel from transmit antenna $m_t$ to target $k$ to receive antenna $m_R$ can be modeled as \cite{kumari2017ieee}
\begin{align}\label{eqn:ULA}
g_{m_R\hspace{-1pt},\hspace{-1pt}m_T}(\hspace{-1pt}d_k\hspace{-1pt},\hspace{-1pt}\theta_k\hspace{-1pt})\hspace{-3pt}=\hspace{-3pt}\sqrt{G(d_k)}e^{\frac{-j2\pi (2d_k\hspace{-1pt}+\hspace{-1pt}(m_T\hspace{-1pt}+\hspace{-1pt}m_R\hspace{-1pt}-\hspace{-1pt}2)d_{{\rm A}}\sin(\theta_k))}{\mu}}.
\end{align}Here, $\mu$ m denotes the signal wavelength, and
\begin{align}\label{eqn:path loss}
G(d_k)=\frac{\mu^2\sigma_{{\rm RCS}}}{64\pi^3 d_k^4},
\end{align}denotes the attenuation caused by the propagation from radar to the target and then from the target to radar as well as the scattering process, where $\sigma_{{\rm RCS}}$ in ${\rm m}^2$ denotes the radar cross section (RCS).

Due to the unique mapping between scattered channels and target location, the radar can first estimate the scattered channels of various targets, based on which the targets' location can then be recovered. In the following, we introduce the radar transmit and receive signal model for estimating the targets' channels and then their distance and angle to the radar.

In the baseband domain, define
\begin{align}\label{eqn:information signal}
\mv{x}_{m_T}=\sqrt{p}[x_{m_T,1},\ldots,x_{m_T,N}]^T=\sqrt{p}\mv{W}^H\mv{s}_{m_T}, \forall m_T,
\end{align}as the time-domain signal for transmit antenna $m_T$ over one OFDM symbol consisting of $N$ OFDM samples, where $p$ denotes the transmit power of the radar, $\mv{W}\in \mathbb{C}^{N\times N}$ is the discrete Fourier transform (DFT) matrix, and $\mv{s}_{m_T}=[s_{m_T,1},\ldots,s_{m_T,N}]^T$ denotes the frequency-domain samples at transmit antenna $m_T$. In practice, the $N$ frequency-domain samples $s_{m_T,1},\ldots,s_{m_T,N}$ consist of the pilot samples and data samples. For convenience, suppose that the first $N_{{\rm P}}$ samples $\mv{s}_{{\rm P},m_T}=[s_{m_T,1},\ldots,s_{m_T,N_{{\rm P}}}]^T$ are used as pilot, while the remaining $N_D=N-N_P$ samples $\mv{s}_{{\rm D},m_T}=[s_{m_T,N_{{\rm P}}+1},\ldots,s_{m_T,N}]^T$ are used as data. In this paper, we assume that only the pilot samples $\mv{s}_{{\rm P},m_T}$ are used by the radar for detecting the targets, because pilot sequences belonging to different transmit antennas can be designed to be orthogonal to each other, i.e., $\mv{s}_{{\rm P},m_T}^H\mv{s}_{{\rm P},m_T'}=0$ if $ m_T\neq m_T'$, when $N_{{\rm P}}\geq M_T$, which is desirable for radar sensing.

At the beginning of each OFDM symbol, a cyclic prefix (CP) consisting of $Q<N$ samples is inserted such that both the radar receiver and information receivers can cancel the inter-symbol interference. As a result, for each OFDM symbol, the transmit signal of transmit antenna $m_T$, $m_T=1,\ldots,M_T$, in the time domain is
\begin{align}\label{eqn:transmit signal}
\bar{\mv{x}}_{m_T}&\hspace{-3pt}=\hspace{-3pt}[\underbrace{\bar{x}_{m_T,-Q},\ldots,\bar{x}_{m_T,-1}}\limits_{{\rm CP}},\underbrace{\bar{x}_{m_T,0},\ldots,\bar{x}_{m_T,N-1}}\limits_{{\rm data}}]^T \nonumber \\ &\hspace{-3pt}=\hspace{-3pt}[\underbrace{x_{m_T,N-Q+1},\ldots,x_{m_T,N}}\limits_{{\rm CP}},\underbrace{x_{m_T,1},\ldots,x_{m_T,N}}\limits_{{\rm data}}]^T,
\end{align}where the first $Q$ samples $\bar{x}_{m_T,-Q},\ldots,\bar{x}_{m_T,-1}$ are the CP.

If target $k$ and target $k'$ are in the same cluster, i.e., $\frac{lc_0}{2N\Delta f}<d_k,d_{k'}\leq \frac{(l+1)c_0}{2N\Delta f}$ for some $l\leq L_{{\rm max}}$, then the scattered signals from them will arrive at the radar receiver within the same sampling period; otherwise, their scattered signals will arrive at the radar receiver in different sampling periods. As a result, in the time domain, the received signal at receive antenna $m_R$ for the $n$th sampling period (after removing the first $Q$ samples corrupted by the CP) is
\begin{align}\label{eqn:received signal 1}
\hspace{-3pt}y_{m_R,n}&\hspace{-3pt}=\hspace{-3pt}\sum\limits_{m_T=1}^{M_T}\hspace{-2pt}\sum\limits_{l=1}^{L_{{\rm max}}}\hspace{-2pt}\sum\limits_{i=1}^{K_l}g_{m_R,m_T}(D_{l,i},\hspace{-2pt}\Theta_{l,i})\sqrt{p}\bar{x}_{m_T,n-l}\hspace{-2pt}+\hspace{-2pt}z_{m_R,n} \nonumber \\
& \hspace{-3pt}=\hspace{-3pt} \sum\limits_{m_T=1}^{M_T}\hspace{-2pt}\sum\limits_{l=1}^{L_{{\rm max}}} h_{m_R,m_T,l} \sqrt{p}\bar{x}_{m_T,n-l}\hspace{-2pt}+\hspace{-2pt}z_{m_R,n}, ~ \forall n.
\end{align}In (\ref{eqn:received signal 1}), $\bar{x}_{m_T,n}$'s are defined in (\ref{eqn:transmit signal}), $z_{m_R,n}\in \mathcal{CN}(0,\sigma^2)$ denotes the superposition of additive white Gaussian noise (AWGN) as well as the scattered signals from the objects out of the radar sensing range ($d_{{\rm max}}$ m), which is assumed to be independent over $m_R$ and $n$, and
\begin{align}\label{eqn:channel 2}
\hspace{-3pt} h_{m_R,m_T,l}=\left\{\begin{array}{ll}\sum\limits_{i=1}^{K_l}g_{m_R,m_T}(D_{l,i},\Theta_{l,i}), & {\rm if} ~ \Omega_l\neq \emptyset, \\ 0, & {\rm otherwise},\end{array}\right.
\end{align}denotes the effective scattered channel of all the targets whose scattering delay is of a duration with $l$ OFDM samples. Note that if $\Omega_l\neq \emptyset$, then $h_{m_R,m_T,l}$ is a function of $d_k$'s and $\theta_k$'s, $\forall k\in \Omega_l$; while if $\Omega_l=\emptyset$, then $h_{m_R,m_T,l}=0$.

\begin{figure*}[t]
\setcounter{equation}{13}
\begin{align}\label{eqn:channel 1}
\bar{\mv{H}}_{m_R,m_T}=\left[\begin{array}{ccccccc} h_{m_R,m_T,1} & 0 & \cdots & 0 & h_{m_R,m_T,L_{{\rm max}}} & \cdots & h_{m_R,m_T,2} \\
h_{m_R,m_T,2} & h_{m_R,m_T,1} & \cdots & 0 & 0 & \cdots & h_{m_R,m_T,3} \\
\vdots & \vdots & \vdots & \vdots & \vdots & \vdots & \vdots \\
0 & 0 & \cdots & h_{m_R,m_T,L_{{\rm max}}} & h_{m_R,m_T,L_{{\rm max}}-1} & \cdots & h_{m_R,m_T,1} \end{array}\right]\in \mathbb{C}^{N\times N},
\end{align}
\hrulefill
\end{figure*}

Over one OFDM symbol consisting of $N$ OFDM samples, the overall received signal at receive antenna $m_R$ is
\begin{align}\setcounter{equation}{12}
\mv{y}_{m_R}&=[y_{m_R,1},\ldots,y_{m_R,N}]^T \nonumber \\ &=\sqrt{p}\sum\limits_{m_T=1}^{M_T}\bar{\mv{H}}_{m_R,m_T}\mv{x}_{m_T}+\mv{z}_{m_R}, ~~~ \forall m_R,
\end{align}where $\mv{x}_{m_T}$ is defined in (\ref{eqn:information signal}), $\mv{z}_{m_R}\in \mathbb{C}^{N\times 1}\sim \mathcal{CN}(\mv{0},\sigma^2\mv{I})$, and $\bar{\mv{H}}_{m_R,m_T}$ is given in (\ref{eqn:channel 1}) on the top of next page. Because each $\bar{\mv{H}}_{m_R,m_T}$ is a circulant matrix in which each row is rotated one element to the right relative to the preceding row, its eigenvalue decomposition (EVD) can be expressed as
\begin{align}\label{eqn:EVD of channel}\setcounter{equation}{14}
\bar{\mv{H}}_{m_R,m_T}=\mv{W}^H\mv{\Lambda}_{m_R,m_T}\mv{W}, ~~~ \forall m_R,m_T,
\end{align}where $\mv{\Lambda}_{m_R,m_T}={\rm diag}(\lambda_{m_R,m_T,1},\ldots,\lambda_{m_R,m_T,N})$ with the $n$th diagonal element denoted by
\begin{align}\label{eqn:channel in frequency}
\hspace{-3pt}\lambda_{m_R,m_T,n}\hspace{-3pt}=\hspace{-3pt}\sum\limits_{l=1}^{L_{{\rm max}}}h_{m_R,m_T,l}\exp\left(\frac{-j2\pi(n\hspace{-2pt}-\hspace{-2pt}1)(l\hspace{-2pt}-\hspace{-2pt}1)}{N}\right).
\end{align}For simplicity, define $\mv{\lambda}_{m_R,m_T}\hspace{-2pt}=\hspace{-2pt}[\lambda_{m_R,m_T,1},\ldots,\lambda_{m_R,m_T,N}]^T$ and $\mv{h}_{m_R,m_T}=[h_{m_R,m_T,1},\ldots,h_{m_R,m_T,L_{{\rm max}}}]^T$, $\forall m_T,m_R$. Then, we have
\begin{align}\label{eqn:lambda}
\mv{\lambda}_{m_R,m_T}= \mv{E}\mv{h}_{m_R,m_T}, ~~~ \forall m_R,m_T,
\end{align}where $\mv{E}\in \mathbb{C}^{N\times L_{{\rm max}}}$ with the entry on the $n$th row and $l$th column denoted by $E_{n,l}=\exp(-j2\pi(n-1)(l-1)/N)$.

In the frequency domain, the radar will apply DFT to the time-domain signals received at each received antenna, i.e.,
\begin{align}\label{eqn:DFT}
\bar{\mv{y}}_{m_R}&=[\bar{y}_{m_R,1},\ldots,\bar{y}_{m_R,N}]^T =\mv{W}\mv{y}_{m_R}\nonumber \\
& = \sqrt{p}\sum\limits_{m_T=1}^{M_T} {\rm diag}(\mv{s}_{m_T})\mv{E}\mv{h}_{m_R,m_T} +\bar{\mv{z}}_{m_R}\nonumber \\ & =\sqrt{p}\mv{B}\mv{h}_{m_R}+\bar{\mv{z}}_{m_R}, ~~~ \forall m_R,
\end{align}where $\mv{h}_{m_R}=[\mv{h}_{m_R,1}^T,\ldots,\mv{h}_{m_R,M_T}^T]^T\in \mathbb{C}^{M_T L_{{\rm max}}\times 1}$, $\bar{\mv{z}}_{m_R}=[\bar{z}_{m_R,1},\ldots,\bar{z}_{m_R,N}]^T=\mv{W}\mv{z}_{m_R}\sim \mathcal{CN}(\mv{0},\sigma^2\mv{I})$ since $\mv{W}\mv{W}^H=\mv{I}$, and
\begin{align}\label{eqn:B}
\mv{B}=[{\rm diag}(\mv{s}_1)\mv{E},\ldots,{\rm diag}(\mv{s}_{M_T})\mv{E}]\in \mathbb{C}^{N\times L_{{\rm max}}M_T}.
\end{align}

Last, the overall received signal across all the $M_R$ radar receive antennas and $N$ sub-carriers is
\begin{align}\label{eqn:overall signal}
\bar{\mv{y}}=[\bar{\mv{y}}_1^T,\ldots,\bar{\mv{y}}_{M_R}^T]^T=\sqrt{p}\mv{\Theta}\mv{h}+\bar{\mv{z}}\in \mathbb{C}^{NM_R\times 1},
\end{align}where
\begin{align}
\mv{\Theta}=\left[\begin{array}{cccc}\mv{B} & \cdots & \mv{0} \\ \vdots & \ddots & \vdots \\ \mv{0} & \cdots & \mv{B} \end{array}\right]\in \mathbb{C}^{NM_R\times L_{{\rm max}}M_TM_R},
\end{align}is a block diagonal matrix, $\mv{h}=[\mv{h}_1^T,\ldots,\mv{h}_{M_R}^T]^T\in \mathbb{C}^{L_{{\rm max}}M_TM_R\times 1}$ denotes the overall channel, and $\bar{\mv{z}}=[\bar{\mv{z}}_1^T,\ldots,\bar{\mv{z}}_{M_R}^T]^T\in \mathbb{C}^{NM_R\times 1}$.

As stated at the very beginning of Section \ref{sec:Signal Model}, the radar merely uses the pilot signal $\mv{s}_{{\rm P},m_T}$ for detecting the targets, which occupies the first $N_{{\rm P}}$ sub-carriers. As a result, for the received signal $\bar{\mv{y}}$, we are only interested in
\begin{align}\label{eqn:received pilot}
\bar{\mv{y}}_{{\rm P}}=[\bar{\mv{y}}_{{\rm P},1}^T,\ldots,\bar{\mv{y}}_{{\rm P},M_R}^T]^T,
\end{align}where $\bar{\mv{y}}_{{\rm P},m_R}=[\bar{y}_{m_R,1},\ldots,\bar{y}_{m_R,N_{{\rm P}}}]^T$ consists of the first $N_{{\rm P}}$ samples in $\bar{\mv{y}}_{m_R}$, $\forall m_R$. Define
\begin{align}
& \bar{\mv{B}}_{{\rm P},l}\hspace{-3pt}=\hspace{-3pt}\left[\begin{array}{ccc}s_{1,1}E_{1,l}\hspace{-3pt} & \hspace{-3pt}\cdots \hspace{-3pt}& \hspace{-3pt} s_{M_T,1}E_{1,l} \\ \vdots \hspace{-3pt}&\hspace{-3pt} \ddots \hspace{-3pt}&\hspace{-3pt} \vdots \\ s_{1,N_{{\rm P}}}E_{N_{{\rm P}},l}\hspace{-3pt} & \hspace{-3pt}\cdots \hspace{-3pt} & \hspace{-3pt} s_{M_T,N_{{\rm P}}}E_{N_{{\rm P}},l} \end{array}\right]\hspace{-3pt}\in \hspace{-3pt} \mathbb{C}^{N\times M_T}, \label{eqn:theta 1}\\
& \bar{\mv{\Theta}}_{{\rm P},l}\hspace{-3pt}=\hspace{-3pt}\left[\begin{array}{cccc}\bar{\mv{B}}_{{\rm P},l} \hspace{-3pt}& \hspace{-3pt}\cdots \hspace{-3pt}&\hspace{-3pt} \mv{0} \\ \vdots \hspace{-3pt}&\hspace{-3pt} \ddots \hspace{-3pt}&\hspace{-3pt} \vdots \\ \mv{0} \hspace{-3pt}&\hspace{-3pt} \cdots \hspace{-3pt}&\hspace{-3pt} \bar{\mv{B}}_{{\rm P},l} \end{array}\right]\hspace{-3pt}\in\hspace{-3pt} \mathbb{C}^{N_{{\rm P}}M_R\times M_TM_R}, ~ \forall l. \label{eqn:theta 2}
\end{align}Then, the received signal given in (\ref{eqn:received pilot}) can be expressed as
\begin{align}\label{eqn:overall signal 1}
\bar{\mv{y}}_{{\rm P}}=\sqrt{p}\bar{\mv{\Theta}}_{{\rm P}}\bar{\mv{h}}+\bar{\mv{z}}_{{\rm P}},
\end{align}where
\begin{align}
& \bar{\mv{\Theta}}_{{\rm P}}=[\bar{\mv{\Theta}}_{{\rm P},1},\ldots,\bar{\mv{\Theta}}_{{\rm P},L_{{\rm max}}}]\in \mathbb{C}^{N_{{\rm P}}M_R\times L_{{\rm max}}M_TM_R}, \label{eqn:theta 3} \\
& \bar{\mv{h}}=[\bar{\mv{h}}_1^T,\ldots,\bar{\mv{h}}_{L_{{\rm max}}}^T]^T\in \mathbb{C}^{L_{{\rm max}}M_TM_R\times 1}, \label{eqn:group channel} \\
& \bar{\mv{h}}_l=[h_{1,1,l},\ldots,h_{1,M_T,l},h_{2,1,l},\ldots,h_{2,M_T,l},\ldots,\nonumber \\ & ~~~~~~~ h_{M_R,1,l},\ldots,h_{M_R,M_T,l}]^T\in \mathbb{C}^{M_TM_R\times 1}, ~~~ \forall l, \\
& \bar{\mv{z}}_{{\rm P}}\hspace{-3pt}=\hspace{-3pt}[\bar{z}_{1,1},\ldots,\bar{z}_{1,N_{{\rm P}}},\ldots,\bar{z}_{M_R,1},\ldots,\bar{z}_{M_R,N_{{\rm P}}}]^T\in \mathbb{C}^{M_RN_{{\rm P}}\times 1}.
\end{align}Note that $\bar{\mv{h}}$ is obtained by re-ordering the entries in $\mv{h}$ such that the scattered channels of each cluster are put together, as shown in (\ref{eqn:group channel}). The job of the radar is to detect all the targets based on (\ref{eqn:overall signal 1}).


\section{Two-Stage Target Sensing Framework}
In this section, we propose a two-stage approach to detect the targets based on the MIMO-OFDM technique. Specifically, in the first stage, the radar estimate the OFDM channels $\bar{\mv{h}}_l$'s, $l=1,\ldots,L_{{\rm max}}$, based on the signal model given in (\ref{eqn:overall signal 1}). Note that in (\ref{eqn:overall signal 1}), $\bar{\mv{\Theta}}$ merely depends on the pilot samples as shown in (\ref{eqn:theta 1}), (\ref{eqn:theta 2}), and (\ref{eqn:theta 3}), and is thus perfectly known by the radar. In the second stage, after $h_{m_R,m_T,l}$'s are estimated in Stage I, each target's distance and angle to the radar can be recovered by solving the equations given in (\ref{eqn:channel 2}). Due to the space limitation, in this work we merely focus on the case when the radar employs ULA such that $g_{m_R,m_T}(D_{l,i},\Theta_{l,i})$'s are modeled by (\ref{eqn:ULA}), while the other antenna array models will be considered in our future work. Then, if for some $l$, the estimated channels are $h_{m_R,m_T,l}=0$, $\forall m_R,m_T$, it is concluded that there are no targets in cluster $\Omega_l$; otherwise, according to (\ref{eqn:ULA}) and (\ref{eqn:channel 2}), the radar estimates $D_{l,i}$'s and $\Theta_{l,i}$'s, $\forall i\leq K_l$, by solving the following equations:
\begin{align}\label{eqn:estimation 1}
& \sum\limits_{i=1}^{K_l}\hspace{-2pt}\Gamma(\hspace{-1pt}D_{l,i}\hspace{-1pt})\Upsilon_{m_R,m_T}(\hspace{-1pt}\Theta_{l,i}\hspace{-1pt})\hspace{-2pt}=\hspace{-2pt}h_{m_R,m_T,l},  \forall (\hspace{-1pt}m_T,m_R\hspace{-1pt})\hspace{-2pt}\in\hspace{-2pt} \mathcal{M},
\end{align}where
\begin{align}
& \Gamma(D_{l,i})=\sqrt{G(D_{l,i})}e^{\frac{-j2\pi 2D_{l,i}}{\mu}}, \label{eqn:Gamma}\\
& \Upsilon_{m_R,m_T}(\Theta_{l,i})=e^{\frac{-j2\pi (m_T+m_R-2)d_{{\rm A}}\sin(\Theta_{l,i})}{\mu}}, ~ \forall i\leq K_l, \label{eqn:Upsilon} \\
& \mathcal{M}\hspace{-2pt}=\hspace{-2pt}\{\hspace{-1pt}(\hspace{-1pt}1,1\hspace{-1pt}),\hspace{-1pt}\ldots,(\hspace{-1pt}1,\hspace{-1pt}M_R\hspace{-1pt}),\hspace{-1pt}(\hspace{-1pt}2,\hspace{-1pt}M_R\hspace{-1pt}),\hspace{-1pt}\ldots,\hspace{-1pt}(\hspace{-1pt}M_T,\hspace{-1pt}M_R\hspace{-1pt})\hspace{-1pt}\}, \label{eqn:antenna}
\end{align}Note that if $m_T+m_R=m_T'+m_R'$, then $\Upsilon_{m_R,m_T}(\Theta_{l,i})=\Upsilon_{m_R',m_T'}(\Theta_{l,i})$, $\forall i$, according to (\ref{eqn:Upsilon}). As a result, (\ref{eqn:estimation 1}) only contains $M_T+M_R-1$ distinct equations when $m_T+m_R=2,\ldots,M_T+M_R$, as defined by $\mathcal{M}$.

It is worth noting that if there is one transmit antenna and one receive antenna for the radar, i.e., $M_T=M_R=1$, then in the case when the number of targets is larger than one in cluster $l$, i.e., $K_l>1$, it is impossible to recover the targets' location since the number of unknowns is larger than the number of equations in (\ref{eqn:estimation 1}). This is the range resolution issue (\ref{eqn:range resolution}) in radar systems as reported in \cite{Richards10}. Nevertheless, if the radar is equipped with multiple antennas, i.e., $M_T>1$, $M_R>1$, then the number of equations is probably larger than the number of unknowns in (\ref{eqn:estimation 1}). In this case, the radar is able to accurately estimate the location of all the targets based on (\ref{eqn:estimation 1}), even if multiple targets are in the same range cluster as defined by (\ref{eqn:target1}). As a result, the MIMO technology is very effective to combat the range resolution issue in Stage II of our proposed two-stage sensing approach.

In the following two subsections, we introduce how to estimate the OFDM channels based on (\ref{eqn:overall signal 1}) in Stage I, and how to utilize MIMO radar technique to estimate the location of the targets based on (\ref{eqn:estimation 1}) in Stage II.

\subsection{Stage I: Leveraging Group LASSO for OFDM Channel Estimation}\label{sec:Stage 1}
It can be observed from (\ref{eqn:channel 2}) that if there is no target in cluster $l$, i.e., $\Omega_l=\emptyset$, then $\bar{\mv{h}}_l=\mv{0}$. As a result, block sparsity exists in $\bar{\mv{h}}$, which motivates us to employ the group LASSO technique \cite{Yuan06} to estimate the OFDM channels based on (\ref{eqn:overall signal 1}). Particularly, given any $\rho>0$, the group LASSO problem can be formulated as
\begin{align}\label{eqn:problem}
\mathop{\mathrm{minimize}}_{\bar{\mv{h}}} ~~~  0.5\|\bar{\mv{y}}_{{\rm P}}-\sqrt{p}\bar{\mv{\Theta}}_{{\rm P}}\bar{\mv{h}}\|_F^2+\rho\sum\limits_{l=1}^{L_{{\rm max}}}\|\bar{\mv{h}}_l\|_2,
\end{align}which is a convex problem, and thus can be solved by CVX \cite{CVX}. In the above problem, a penalty term $\sum_{l=1}^{L_{{\rm max}}}\|\bar{\mv{h}}_l\|_2$ is present to encourage the block sparsity in $\bar{\mv{h}}$. Moreover, $\rho$ can balance between the estimation error and sparsity in $\bar{\mv{h}}$: as $\rho$ increases, more blocks in $\bar{\mv{h}}$ tend to be zero, but the term of the estimation error in the objective function plays a smaller role. In practice, we can keep increasing the value of $\rho$ until the resulting solution to problem (\ref{eqn:problem}) achieves a good trade-off between the estimation error and channel sparsity.

\subsection{Stage II: Leveraging MIMO for Detecting Targets in the Same Range Cluster}\label{sec:Stage 2}
After the OFDM channels $\bar{\mv{h}}_l$'s are estimated in Stage I, in this section we introduce how to estimate the targets' location by solving the equations given by (\ref{eqn:estimation 1}) in Stage II.

Note that in (\ref{eqn:estimation 1}), besides $D_{l,i}$'s and $\Theta_{l,i}$'s, the numbers of targets in various clusters $K_l$'s are also unknown and thus need to be estimated. It is shown in \cite{Stoica07} that in cluster $l$, if the number of targets satisfies
\begin{align}\label{eqn:upper bound}
K_l\leq K_{{\rm max}}\triangleq \left \lfloor \frac{M_T+M_R-2}{2} \right \rfloor,
\end{align}then, there is always a unique solution of $D_{l,i}$'s and $\Theta_{l,i}$'s to the equations given in (\ref{eqn:estimation 1}). In this work, we assume that condition (\ref{eqn:upper bound}) is true for all the clusters. Then, for any $l\leq L_{{\rm max}}$, if $h_{m_R,m_T,l}=0$, $\forall m_R,m_T$, it is concluded that there are no targets in cluster $l$, i.e., $K_l^\ast=0$; otherwise, we do exhaustive search to find the number of targets in cluster $l$ in the regime of $[1,K_{{\rm max}}]$. Particularly, given any $\bar{K}_l\in [1,K_{{\rm max}}]$, the corresponding target location is estimated by solving the following sub-problem:
\begin{align*}&(\mathrm{P}-\bar{K}_l):\\
&\mathop{\mathrm{minimize}}_{\{\hspace{-1pt}D_{l,i},\Theta_{k,i}\hspace{-1pt}\}}  \hspace{-2pt} \sum\limits_{(m_T,m_R)\in \mathcal{M}}\hspace{-2pt}\left\|\sum\limits_{i=1}^{\bar{K}_l}\hspace{-1pt}\Gamma(\hspace{-1pt}D_{l,i}\hspace{-1pt})\Upsilon_{m_R,m_T}(\hspace{-1pt}\Theta_{l,i}\hspace{-1pt})\hspace{-2pt}-\hspace{-2pt}h_{m_R,m_T,l}\right\|_2^2
\end{align*}where $\mathcal{M}$ is defined in (\ref{eqn:antenna}) and $h_{m_R,m_T,l}$'s are estimated in Stage I.

Let $\Phi(\bar{K}_l)$ denote the optimal value of problem (P-$\bar{K}_l$), $\forall \bar{K}_l$. Then, the number of targets in cluster $l$ is detected as
\begin{align}\label{eqn:optimal number}
K_l^\star=\arg\min\limits_{1\leq \bar{K}_l \leq K_{{\rm max}}} \Phi(\bar{K}_l), ~ {\rm if} ~ \bar{\mv{h}}_l\neq \mv{0}.
\end{align}Moreover, let $D_{l,i}^\star$'s and $\Theta_{l,i}^\star$'s, $i=1,\ldots,K_l^\star$, denote the optimal solution to problem (P-$K_l^\star$). Then, the location of the targets in cluster $l$ can be determined by $D_{l,i}^\star$'s and $\Theta_{l,i}^\star$'s.

The main challenge thus lies in how to solve problem (P-$\bar{K}_l$) given some $\bar{K}_l\in [1,K_{{\rm max}}]$. In this work, we propose to quantize each $\Theta_{l,i}$ in the regime $(0,\theta_{{\rm max}}]$, i.e., $\bar{\Theta}_{l,i}=\Delta \theta,2\Delta \theta,\ldots,\theta_{{\rm max}}$, where $\Delta \theta$ is the size of the quantization interval. Given each combination of the quantized angle denoted by $\bar{\mv{\theta}}_l=[\bar{\Theta}_{l,1},\ldots,\bar{\Theta}_{l,\bar{K}_l}]^T$, problem (P-$\bar{K}_l$) reduces to
\begin{align*}&(\mathrm{P}-\bar{K}_l-\bar{\mv{\theta}}_l):\\
&\mathop{\mathrm{minimize}}_{\{\Gamma(D_{l,i})\}} \hspace{-2pt}  \sum\limits_{(m_T,m_R)\in \mathcal{M}}\hspace{-2pt}\left\|\sum\limits_{i=1}^{\bar{K}_l}\hspace{-2pt}\Gamma(\hspace{-1pt}D_{l,i}\hspace{-1pt})\Upsilon_{m_R,m_T}(\hspace{-1pt}\bar{\Theta}_{l,i}\hspace{-1pt})\hspace{-2pt}-\hspace{-2pt}h_{m_R,m_T,l}\right\|_2^2
\end{align*}Note that the optimization variables are changed from $D_{l,i}$'s to $\Gamma
(D_{l,i})$'s due to the one-to-one correspondence between $D_{l,i}$'s and $\Gamma(D_{l,i})$'s as shown in (\ref{eqn:Gamma}). Problem (P-$\bar{K}_l$-$\bar{\mv{\theta}}_l$) is a convex problem over $\Gamma(D_{l,i})$'s, and it can be shown that if (\ref{eqn:upper bound}) is true, there is a unique optimal solution denoted by
\begin{align}\label{eqn:optimal distance}
\hspace{-5pt}[\bar{\Gamma}(\hspace{-1pt}D_{l,1}\hspace{-1pt}),\hspace{-1pt}\ldots,\hspace{-1pt}\bar{\Gamma}(\hspace{-1pt}D_{l,\bar{K}_l}\hspace{-1pt})]^T\hspace{-4pt}=\hspace{-4pt}\left(\hspace{-2pt}\mv{\Upsilon}_{\mathcal{M}}(\hspace{-1pt}\bar{\mv{\theta}}_l\hspace{-1pt})\mv{\Upsilon}_{\mathcal{M}}^H(\hspace{-1pt}\bar{\mv{\theta}}_l\hspace{-1pt})\hspace{-2pt}\right)^{-1}\hspace{-3pt}\mv{\Upsilon}_{\mathcal{M}}(\hspace{-1pt}\bar{\mv{\theta}}_l\hspace{-1pt})\mv{h}_{\mathcal{M},l},
\end{align}where $\mv{\Upsilon}_{\mathcal{M}}(\bar{\mv{\theta}}_l)\in \mathbb{C}^{\bar{K}_l\times (M_T+M_R-1)}$ and $\mv{h}_{\mathcal{M},l}\in \mathbb{C}^{(M_T+M_R-1)\times 1}$ are given by
\begin{align}
& \mv{\Upsilon}_{\mathcal{M}}(\bar{\mv{\theta}}_l)=[\ldots,\mv{\Upsilon}_{m_R,m_T}(\hspace{-1pt}\bar{\mv{\theta}}_l\hspace{-1pt}),\ldots]_{\forall (m_T,m_R)\in \mathcal{M}}, \\
& \mv{\Upsilon}_{m_R,m_T}(\hspace{-1pt}\bar{\mv{\theta}}_l\hspace{-1pt})\hspace{-2pt}=\hspace{-2pt}[\Upsilon_{m_R,m_T}^\ast(\hspace{-1pt}\bar{\Theta}_{l,1}\hspace{-1pt}),\ldots,\Upsilon_{m_R,m_T}^\ast(\hspace{-1pt}\bar{\Theta}_{l,\bar{K}_l}\hspace{-1pt})]^T, \\
& \mv{h}_{\mathcal{M},l}=[\ldots,h_{m_R,m_T,l},\ldots]_{\forall (m_T,m_R)\in \mathcal{M}}^T.
\end{align}

Let $\bar{\Phi}(\bar{K}_l,\bar{\mv{\theta}}_l)$ denote the optimal value of problem (P-$\bar{K}_l$-$\bar{\mv{\theta}}_l$). Then, it follows that
\begin{align}\label{eqn:sub objective}
\Phi(\bar{K}_l)\hspace{-3pt}=\hspace{-3pt}\min\limits_{\{\bar{\Theta}_{l,i}\hspace{-1pt}=\hspace{-1pt}\Delta \theta,2\Delta \theta,\ldots,\theta_{{\rm max}}\}} \hspace{-3pt} \bar{\Phi}(\bar{K}_l,\bar{\mv{\theta}}_l),  \bar{K}_l\in[1,K_{{\rm max}}].
\end{align}Because the optimal solution to problem (P-$\bar{K}_l$-$\bar{\mv{\theta}}_l$) can be characterized in closed form as given in (\ref{eqn:optimal distance}), the complexity to obtain $\Phi(\bar{K}_l)$ based on exhaustive search over the quantized angle is moderate.

\begin{remark}\label{remark1}
It can be observed from (\ref{eqn:Upsilon}) that if $\theta_{{\rm max}}>\pi/2$, then we have $\Upsilon_{m_R,m_T}(\Theta_{l,i})=\Upsilon_{m_R,m_T}(\Theta_{l,i}+\pi/2)$. In this case, the estimated angle from each target to the radar based on (\ref{eqn:sub objective}) is not unique. As a result, under the proposed sensing scheme, the maximum sensing angle should be limited as
\begin{align}
\theta_{{\rm max}}\leq \pi/2.
\end{align}Moreover, even with $\theta_{{\rm max}}\leq \pi/2$, if the antenna spacing satisfies $d_{{\rm A}}\sin(\theta_{{\rm max}})\geq \mu$, it is still possible that $\Upsilon_{m_R,m_T}(\Theta_{l,i})=\Upsilon_{m_R,m_T}(\Theta_{l,i}')$ holds for some $0< \Theta_{l,i}\neq \Theta_{l,i}'\leq \theta_{{\rm max}}$. As a result, the antenna spacing should satisfy
\begin{align}
d_{{\rm A}}\leq \frac{\mu}{\sin(\theta_{{\rm max}})}.
\end{align}
\end{remark}

Last, consider problems (P-$K_l^\star$-$\bar{\mv{\theta}}_l$)'s, for $\bar{\Theta}_{l,i}=\Delta \theta,2\Delta \theta,\ldots,\theta_{{\rm max}}$ and $k_l=1,\ldots,K_l^\star$. Define
\begin{align}\label{eqn:optimal angle}
\hspace{-5pt}\mv{\theta}_l^\star\hspace{-3pt}=\hspace{-3pt}[\hspace{-1pt}\Theta_{l,1}^\star\hspace{-1pt},\hspace{-1pt}\ldots,\hspace{-1pt}\Theta_{l,K_l^\star}^\star\hspace{-1pt}]^T\hspace{-3pt}=\hspace{-3pt}\arg\hspace{-3pt}\min\limits_{\{\hspace{-1pt}\bar{\Theta}_{l,i}=\Delta \theta\hspace{-1pt},2\Delta \theta\hspace{-1pt},\ldots,\theta_{{\rm max}}\hspace{-1pt}\}}\hspace{-3pt} \bar{\Phi}(\hspace{-1pt}K_l^\star\hspace{-1pt},\hspace{-1pt}\bar{\mv{\theta}}_l\hspace{-1pt}),
\end{align}as the optimal angle solution of the $K_l^\star$ targets in cluster $l$. Moreover, define $\Gamma_{l,1}^\star,\ldots,\Gamma_{l,K_l^\star}^\star$ as the optimal solution to problem (P-$K_l^\star$-$\mv{\theta}_l^\star$), which can be obtained according to (\ref{eqn:optimal distance}). According to (\ref{eqn:path loss}) and (\ref{eqn:Gamma}), the optimal distance solution of the $K_l^\star$ targets in cluster $l$, denoted by $D_{l,1}^\star,\ldots,D_{l,K_l^\star}^\star$, can thus be obtained via
\begin{align}\label{eqn:range}
D_{l,i}^\star=\sqrt[4]{\frac{\mu^2 \sigma_{{\rm RCS}}}{64\pi^3|\Gamma_{l,i}^\star|^2}}, ~ i=1,\ldots,K_l^\star.
\end{align}

To summarize, the algorithm to estimate the location of all the targets in cluster $l$, $l=1,\ldots,L_{{\rm max}}$, is presented in Algorithm \ref{table2}.

\begin{algorithm}
If $\bar{\mv{h}}_l=\mv{0}$, then the number of targets in cluster $l$ is $0$; otherwise
\begin{itemize}
\item[1] For each $\bar{K}_l\in [1,K_{{\rm max}}]$, solve problems (P-$\bar{K}_l$-$\bar{\mv{\theta}}_l$)'s via (\ref{eqn:optimal distance}) and then obtain the optimal values $\bar{\Phi}(\bar{K}_l,\bar{\mv{\theta}}_l)$'s with $\bar{\Theta}_{l,k_l}=\Delta \theta,2\Delta \theta,\ldots,\theta_{{\rm max}}$, $k_l=1,\ldots,\bar{K}_l$;
\item[2] For each $\bar{K}_l\in [1,K_{{\rm max}}]$, set the optimal value of problem (P-$\bar{K}_l$) as $\Phi(\bar{K}_l)$ according to (\ref{eqn:sub objective});
\item[3] Set the number of targets in cluster $l$, $K_l^\star$, based on (\ref{eqn:optimal number});
\item[4] Set the angle of the $K_l^\star$ targets, $\mv{\theta}_l^\star=[\Theta_{l,1}^\star,\ldots,\Theta_{l,K_l^\star}^\star]^T$, based on (\ref{eqn:optimal angle});
\item[5] Set $\Gamma_{l,i}^\star$'s, $i=1,\ldots,K_l^\star$, as the optimal solution to problem (P-$K_l^\star$-$\mv{\theta}_l^\star$) according to (\ref{eqn:optimal distance}), and set the target range $D_{l,1}^\star,\ldots,D_{l,K_l^\star}^\star$ based on (\ref{eqn:range}).
\end{itemize}
\caption{Proposed Algorithm for Estimating the Number and Location of Targets in Cluster $l$ in Stage II}
\label{table2}
\end{algorithm}

\section{Numerical Results}\label{sec:Numerical Examples}

In this section, we provide one numerical example to verify the performance of our proposed two-stage radar sensing approach. In this example, we assume that the carrier frequency is 2.6 GHz, while $N=1024$ and $\Delta f=15$ KHz such that the range resolution is $\Delta d=9.77$ m. The number of pilot samples is $N_p=300$. Further, it is assumed that $L_{{\rm max}}=10$ and $d_{{\rm max}}=L_{{\rm max}}\Delta d=97.7$ m. According to Remark \ref{remark1}, we set $\theta_{{\rm max}}=\pi/2$ and $d_{{\rm A}}=0.9\mu$. We assume there are $K=4$ targets, while targets 1 and 2 are in cluster 3, and targets 3 and 4 are in cluster 9. The location of each target is shown in Tables \ref{table3} and \ref{table4}, respectively. The RCS is set to be $\sigma_{{\rm RCS}}=7$ dBsm, and the radar transmit power is $33$ dBm. Moreover, the power spectral density of the AWGN at the radar receiver is $-169$dBm/Hz. Last, it is assumed that $M_T=M_R=M$.

Under the above setup, we show the estimated range and angle of these 4 targets achieved by our proposed two-stage radar sensing approach in Tables \ref{table3} and \ref{table4}, respectively, given different values of $M$. It is observed that when $M=1$, i.e., both the radar transmitter and radar receiver are equipped with one antenna, there is a dramatic gap between the estimated range and angle with the real values. This is because for $l=3$ and $l=9$, the numbers of equations and variables in (\ref{eqn:estimation 1}) are 1 and 4, respectively, and thus the corresponding estimation is very poor. When $M=2$, it is observed that the location estimation performance of targets 3 and 4 in cluster 9 is much improved, but the location estimation performance of targets 1 and 2 in cluster 3 is still very poor. Last, if $M=4$, it is observed that the estimated range and angle of all the targets are very close to the real range and angle. As a result, MIMO radar is very powerful for improving the sensing accuracy under our proposed two-stage strategy.

\begin{table}[t]
\centering
\begin{center}
\caption{Range Estimation} \label{table3}
{\small
\begin{tabular}{|c|c|c|c|c|}
\hline Target & 1 & 2 & 3 & 4 \\
\hline
Real Range (m) & 22.765 & 28.170 & 81.611 & 86.623 \\
\hline
Estimation (m): $M=1$ & 38.048 & 38.048 & 100.363 & 100.363 \\
\hline
Estimation (m): $M=2$ & 5.623 & 5.700 & 80.544 & 80.874 \\
\hline
Estimation (m): $M=4$ & 23.036 & 28.618 & 81.657 & 86.402 \\
\hline
\end{tabular}
}
\end{center}
\end{table}

\begin{table}
\centering
\begin{center}
\caption{Angle Estimation} \label{table4}
{\small
\begin{tabular}{|c|c|c|c|c|}
\hline Target & 1 & 2 & 3 & 4 \\
\hline
Real Angle & $78.810^\circ$ & $83.228^\circ$ & $31.903^\circ$ & $10.404^\circ$ \\
\hline
Estimation: $M=1$ & $0.25^\circ$ & $0.25^\circ$ & $0.25^\circ$ & $0.25^\circ$ \\
\hline
Estimation: $M=2$ & $81.500^\circ$ & $81.750^\circ$ & $31.500^\circ$ & $10.750^\circ$ \\
\hline
Estimation: $M=4$ & $78.750^\circ$ & $83.250^\circ$ & $31.750^\circ$ & $10.500^\circ$ \\
\hline
\end{tabular}
}
\end{center}\vspace{-10pt}
\end{table}

\section{Conclusions}\label{sec:Conclusions}
This paper proposed a two-stage signal processing approach to sense the environment using the MIMO-OFDM based communication signals. In the first stage, the radar estimates the scattered channels from the radar transmitter to various targets to the radar receiver based on the techniques of OFDM channel training and compressed sensing. In the second stage, based on the MIMO radar technique, the radar extracts the location information of these targets from the estimated channels even if some of the targets are in the same range cluster. Our proposed radar sensing strategy is compatible with the legacy MIMO-OFDM based cellular systems, and achieves accurate sensing according to the numerical results.

\bibliographystyle{IEEEtran}
\bibliography{CIC}

\end{document}